\documentclass[epjCONF,columns]{svjour} 
\usepackage{graphics}
\usepackage[varg]{txfonts} 
\usepackage[latin1]{inputenc}
\session-title{Hadron Collider Physics Symposium 2011 - HCP 2011}
\begin{document}
\title{Results from LHCf Experiment}
\author{Alessia Tricomi on behalf of the LHCf Collaboration}
\institute{University of Catania and INFN Catania, Italy}
\abstract{The LHCf experiment has taken data in 2009 and 2010 p-p collisions at LHC at $\sqrt{s} = 0.9$ TeV and $\sqrt{s} = 7$ TeV. The measurement of the forward neutral particle spectra 
produced in proton-proton collisions at LHC up to an energy of 14 TeV in the center of mass system are of  
fundamental importance to calibrate the Monte Carlo models widely used in the 
high energy cosmic ray (HECR) field, up to an equivalent laboratory energy of the order of $10^{17}$ eV. \\
In this paper the first results on the inclusive photon spectrum measured by LHCf is reported. 
Comparison of this spectrum with the model expectations show significant discrepancies, 
mainly in the high energy region. \\
In addition, perspectives for future analyses as well as the 
program for the next data taking period, in particular the possibility to take data in p-Pb collisions, will be discussed.
} 
\maketitle
\section{Introduction}
\label{intro}
The LHCf experiment at LHC has been designed to calibrate the hadron interaction 
models used in High Energy Cosmic Ray (HECR) Physics through the measurement of the forward neutral 
particle produced in p-p interactions. The uncertainty caused by the poor knowledge of the interaction
between very high energy primary cosmic rays and the earth's atmosphere prevents the precise 
deduction of astrophysical parameters from the observational data, thus 
causing the most important source of systematic error in HECR dedicated experiments. 
 Dedicated extensive air shower experiments are in place since many years and have 
strongly contributed to our understanding of High and Ultra High Energy Cosmic (UHECR) 
Ray Physics. Recently, in particular, the Pierre Auger Collaboration~\cite{Abreu:2011pj} and 
the Telescope Array Collaboration~\cite{Tsunesada:2011mp}, thanks to the excellent performance of their 
hybrid detector arrays, are providing us new exciting observations of UHECRs. Although these 
recent results have brought a deeper insight in primary cosmic ray properties, still they 
are largely affected by the poor knowledge of the nuclear interactions   
in the earth's atmosphere. 
A calibration of the energy scale in the $10^{15}\div 10^{17}$ eV energy range accessible to
 LHC provides crucial input for a better interpretation of primary cosmic ray properties, 
in the region between the ``knee'' and the GZK cut-off.

\section{The LHCf experiment}
\label{sec:1}
The LHCf experiment is composed by two independent position sensitive electromagnetic calorimeters, 
located on both side of the ATLAS experiment, 140 m away from the LHC-IP1 interaction point,  
inside the zero-degree neutral absorber 
(Target Neutral Absorber, TAN). Charged particles from the IP are swept away by the inner beam separation dipole 
before reaching the TAN, so that only photons mainly from $\pi^0$ decays, neutrons and 
neutral kaons reach the LHCf calorimeters. \\
Each calorimeter (ARM1 and ARM2) has a double tower structure, with the smaller tower located 
at zero degree collision angle, approximately 
covering the region with pseudo-ra\-pi\-di\-ty $\eta > 10$ and the larger one, approximately covering the region 
with $8.4 < \eta < 10$. Four X-Y layers of position sensitive detectors 
(scintillating fibers in ARM1, silicon micro-strip detectors in ARM2) 
provide measurements of the transverse profile of the showers. The two tower 
structure allows to reconstruct the $\pi^0$ decaying in two $\gamma$s, hitting separately the two towers, 
hence providing a very precise absolute energy calibration of the detectors. In the range E$> 100$ GeV, 
the LHCf detectors have energy and position resolutions for electromagnetic showers better than 5\% and $200 \mu$m, 
respectively. A detailed description of the LHCf experimental set-up and of the expected physics 
performances can be found in Ref.~\cite{jinst}.

\section{The single photon energy spectrum}
\label{sec:2}
The LHCf Collaboration has recently published the measurement of the single photon energy spectrum at 
7 TeV p-p collisions in two pseudo-rapidity bins~\cite{plb}. 
For this analysis only a small subset of the 7 TeV p-p collision data has been used, corresponding 
to an integrated luminosity of 0.68 nb$^{-1}$ and 0.52 nb$^{-1}$ for the ARM1 and ARM2 detectors, respectively. 
The analysed data have been chosen in a particularly clean and low luminosity fill, to minimize backgrounds, 
hence reducing the systematics of the measurement.

The main steps of the analysis work-flow with special emphasis to possible implication 
for the calibration of Monte Carlo models used in HECR Physics are summarised in the following. 

The energy of photons is reconstructed from the signal released by the shower particles in the scintillators,
after applying corrections for the non-uniformity of light collection and for particles 
leaking in and out of the edges of the calorimeter towers. In order to correct for these last two effects, which are 
rather important due to the limited transverse size of both the calorimetric towers, the transverse 
impact position of showers provided by the position sensitive detectors is used.  

\begin{figure}
\begin{center}
\resizebox{0.99\columnwidth}{!}{\includegraphics{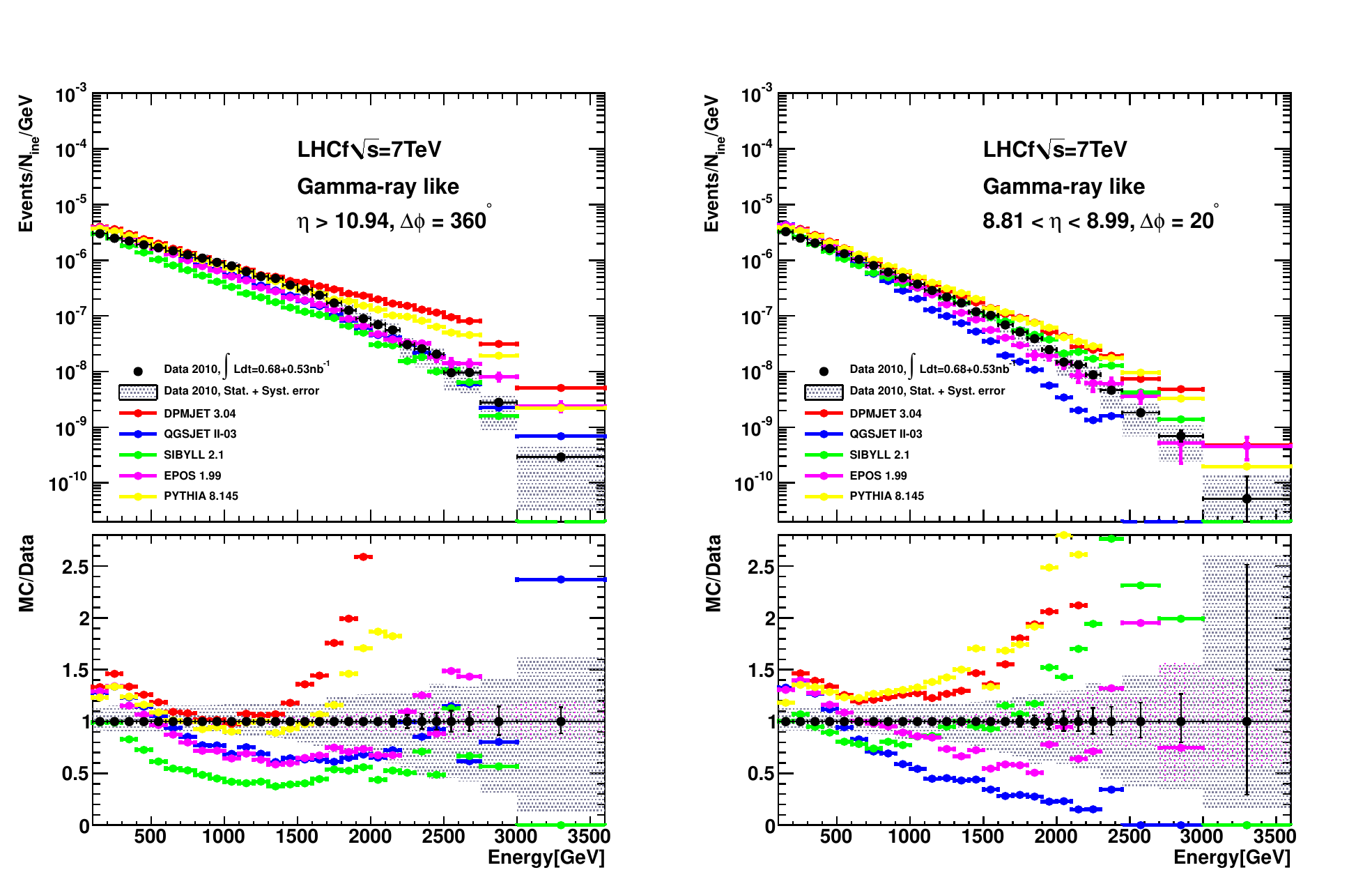}}
\caption{Single photon energy spectra measured by LHCf (black dots) for $\eta > 10.94$ (left) and $8.81 < \eta < 8.99$ (right)  compared to the predictions of DPMJET III~3.04 (red), QGSJET II-03 (blue), SIBYLL 2.1 (green), EPOS 1.99 (magenta) and PYTHIA 8.145 (yellow). Top panels show the spectra and bottom panels show the ratio of MC 
results to experimental data. Error bars and gray shaded areas in each plot indicate the statistical and the 
systematic errors, respectively. The blue shaded area indicates the statistical error of the MC data. Figure from 
Ref.~\cite{plb}.}
\label{fig.spectra}
\end{center}
\end{figure}

Event produced by neutral hadrons are rejected using the information about the longitudinal 
development of the showers, which is different for electromagnetic and hadronic particles.
In addition, thanks to the information provided by the position sensitive detectors, 
events with more than one shower inside the same tower (multi-hit) are rejected. 
In order to combine the spectra measured by ARM1 and ARM2, which have different geometrical 
configurations, in this analysis only events detected in a common pseudo-rapidity and azimuthal range are selected: 
$\eta > 10.94$ and $\Delta\phi = 360^{\mathrm o}$ for the small towers and $8.81 < \eta < 8.99$ and $\Delta\phi = 20^{\mathrm o}$ 
for the large towers.
Figure~\ref{fig.spectra} shows the single $\gamma$ spectrum measured by LHCf in the two pseudo-rapidity regions 
compared with results predicted by MC simulations using different models: 
DPMJET III-3.04~\cite{DPMJET}, QGSJET II-03~\cite{qgsjet}, SIBYLL 2.1~\cite{sybill}, EPOS 1.9~\cite{epos} and 
PYTHIA 8.145~\cite{pythia}. Statical errors and systematic uncertainties are also plotted. A careful study of 
systematic uncertainties has been done and conservative estimates have been taken into account. 
Further details can be found in Ref.~\cite{plb}. As can be seen from Fig.~\ref{fig.spectra}, a clear discrepancy 
between the experimental results and the predictions of the models in the whole energy region is present. 

\section{Impact of LHCf results on HECR Physics}
\label{sec:3}

The first LHCf results on the photon energy spectra have raised attention in the HECR community. As reported in 
the previous paragraph, none of the models agree in the whole energy range with the data. Tuning of the models are 
hence needed to describe the Physics of hadronic interactions at the TeV scale.
In order to better understand the implication of this measurement for the HECR Physics, a collaboration with 
several MC developers and theoreticians has started.

\begin{figure}
\begin{center}
\resizebox{0.99\columnwidth}{!}{\includegraphics{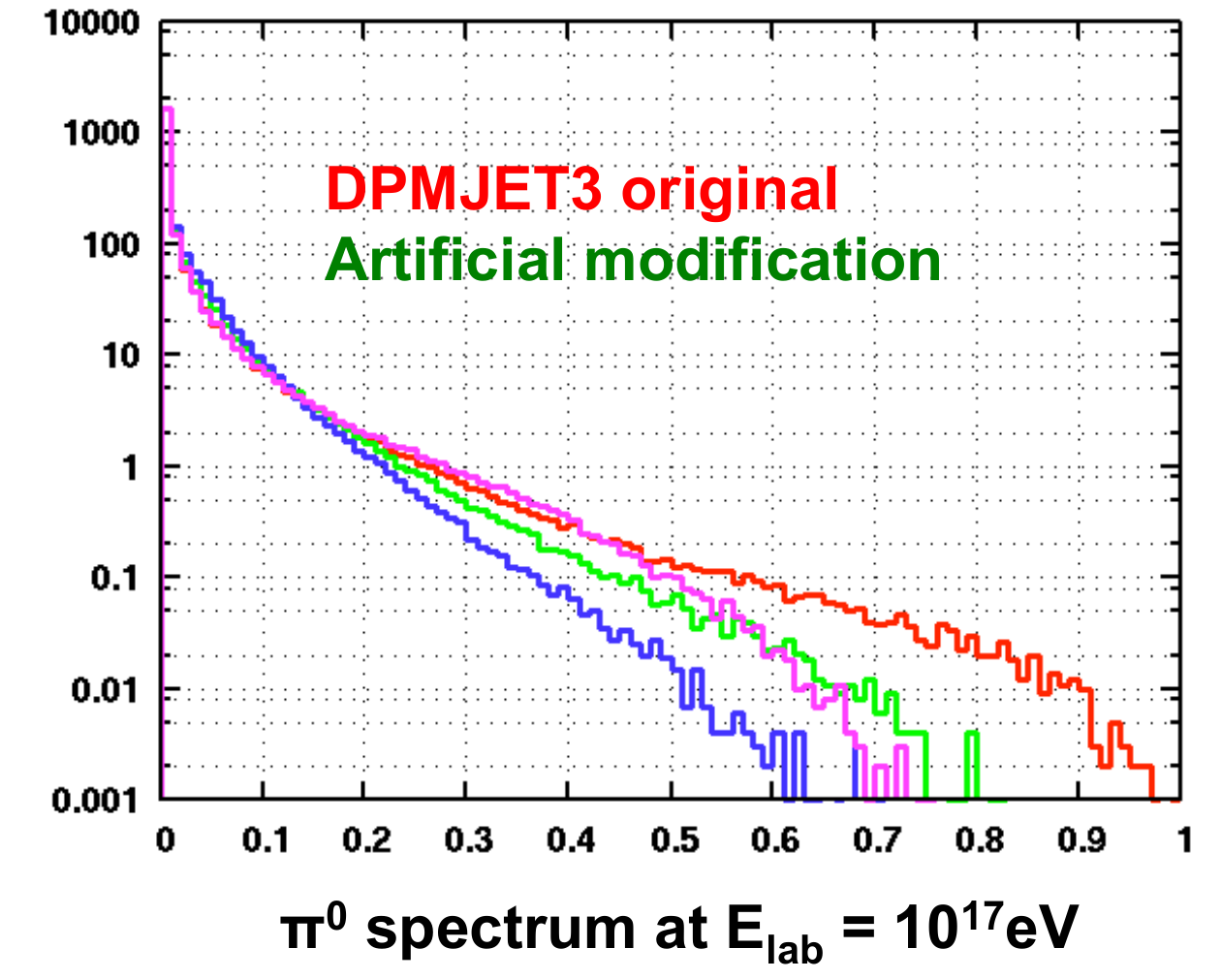}}
\resizebox{0.99\columnwidth}{!}{\includegraphics{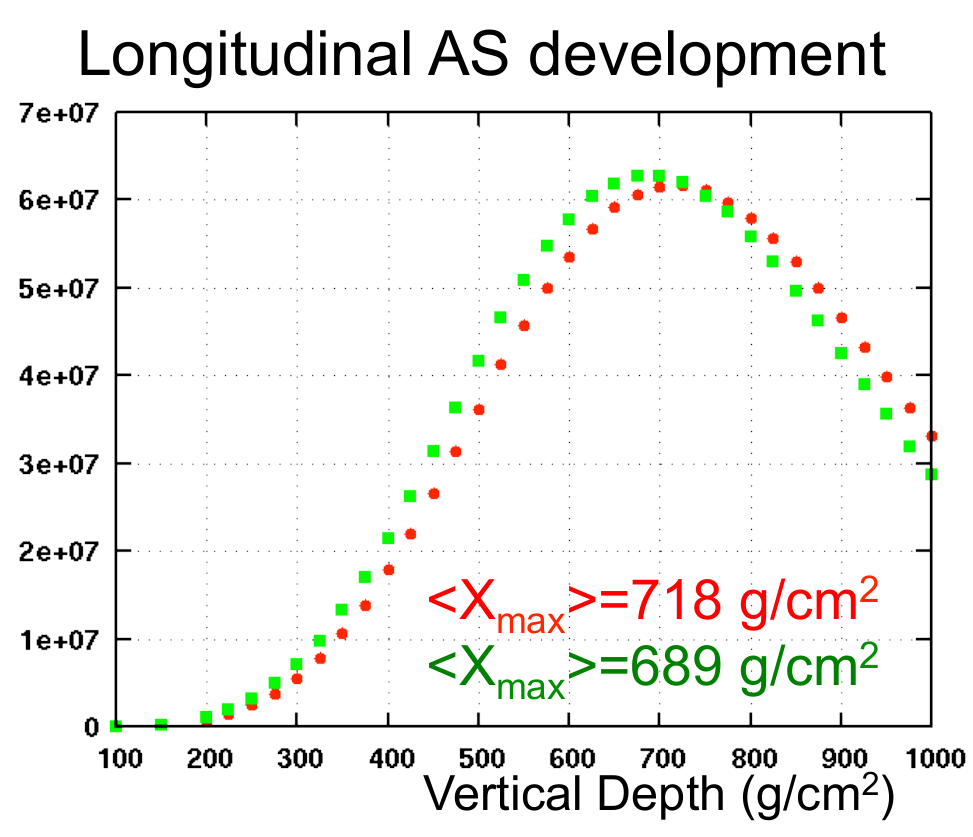}}
\caption{$\pi^0$ spectra at E$_{lab}=10^{17}$ eV for original DPMJET III-3.04 model (red) and artificially modified models (gree, blue, magenta), see text (top). Average longitudinal development of the atmospheric shower in the DPMJET III-3.04 model (red points) and in the artificially modified model (green points) (bottom).}
\label{fig.kasahara}
\end{center}
\end{figure}

As an example, we have artificially modified the DPMJET III~3.04 model to produce a $\pi^0$ spectrum that 
differs from the original one by an amount approximately equal to the difference 
expected between the different models.
Figure~\ref{fig.kasahara} (top) shows the $\pi^0$ spectra at E$_{lab}=10^{17}$ eV predicted by the original 
  DPMJET III~3.04 model and the artificially modified ones, while on the bottom panel  
the average longitudinal development of the atmospheric shower in the original 
model (red points) and in the artificially modified model (green points) is shown.
A difference in the position of the shower maximum of the order of 30 g/cm$^2$ is observed.  

\begin{figure}
\begin{center}
\resizebox{0.99\columnwidth}{!}{\includegraphics{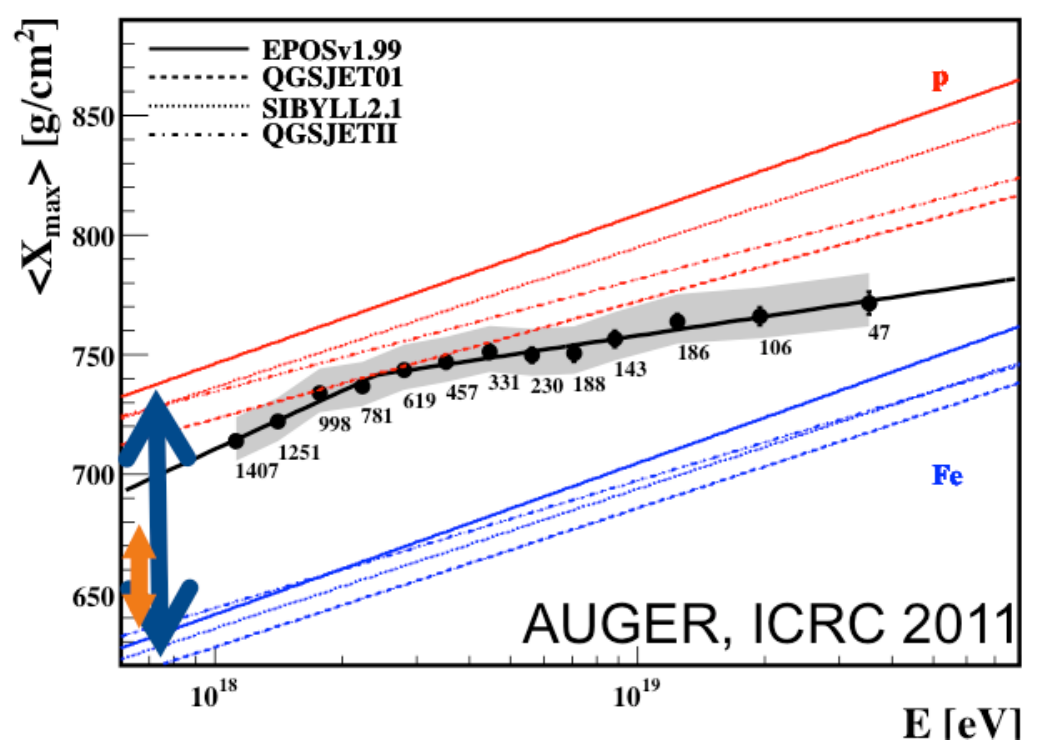}}
\caption{$<{\mathrm X}_{max}>$ distribution as measured by Auger~\protect{\cite{auger}} (black points) 
compared with the model expectations for a light (red) 
or heavy (blue) cosmic ray composition. The yellow arrow correspond to the 30 g/cm$^2$ shift, obtained in Fig.~2.}
\label{fig.composition}
\end{center}
\end{figure}

Fig.~\ref{fig.composition} shows, as an example, the most recent results 
by Auger Collaboration~\cite{auger} for the distribution of the \break $<{\mathrm X}_{max}>$ variable as function of the energy, 
which is the most used method to infer cosmic rays composition,  
compared with the model predictions for a proton-like (red lines) and an Iron-like (blue lines) cosmic ray 
components, respectively. The difference in the $<{\mathrm X}_{max}>$ distributions for the two cases is of the order of 
100 g/cm$^2$ hence a 30 g/cm$^2$ shift is a sizable difference which may significantly reflect in the interpretation 
of HECR data. \\
The importance of a direct measurement of the $\gamma$ and $\pi^0$ spectra by LHCf is clear. \\

\section{Perspective for a Proton/Lead run at LHC}
\label{sec:4}
The data gathered by LHCf both at 900 GeV and 7 TeV center of mass energy are extremely useful for the calibration of hadronic interaction models used for the study of the development of atmospheric showers. Anyway the study of the very forward interactions in the pp system does not allow to have a complete picture of hadronic interactions in atmosphere, where mainly nitrogen and oxygen nuclei are involved, since the effect of the medium can not be neglected if precise estimations are requested. 
The LHCf original physics performances can hence be nicely improved by including measurements of the particles 
very forward produced in the proton-Nuclei (pA) collisions. In principle the ideal way would be to perform  the measurement 
of proton interactions with nitrogen or oxygen nuclei. This is not possible at the moment neither at the LHC nor at other facilities, at least for TeV energies. The LHC though, has instead the unique possibility to study the nuclear effects in the forward region of the collisions between protons and Lead nuclei at the TeV scale, corresponding to a proton energy in the LAB frame exceeding 10$^{16}$ eV. Previous measurements done at other facilities~\cite{RHIC} reported a marked reduction of the 
production cross section values for low transverse momentum events in the forward region of ion interactions, 
with respect to what could be expected by considering the events as binary collisions between nucleons. These results, 
normally reported as the ratio between cross sections measured in p-ion (or ion-ion) and in pp collisions 
(that is the Nuclear Modification Factor, NMF), are currently explained with nucleon screening effects and 
saturation of partonic densities, but so far we have only a poor knowledge of the mechanisms involved, 
especially at very high rapidity where no measurements currently exist. 

\begin{figure}
\begin{center}
\resizebox{0.99\columnwidth}{!}{\includegraphics{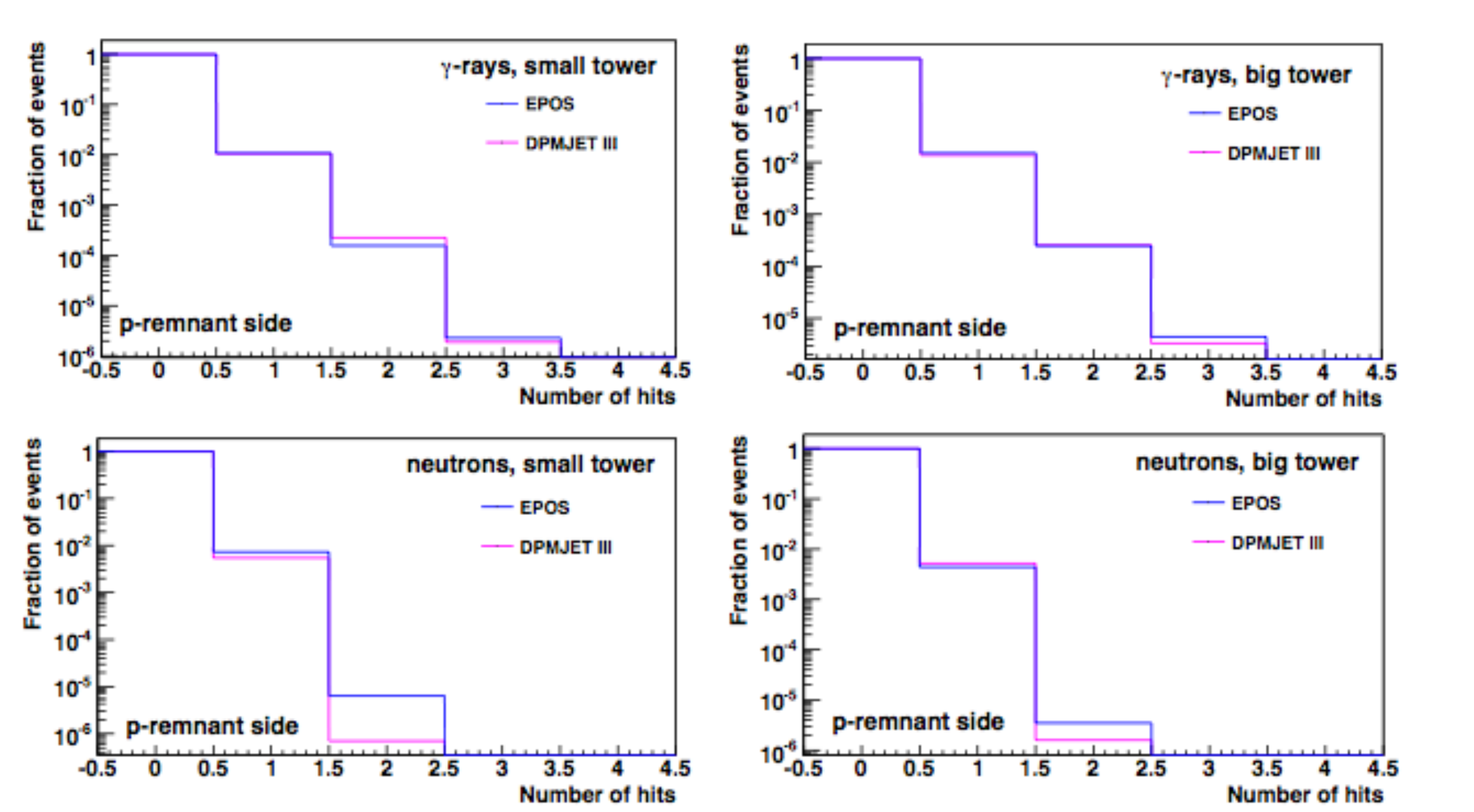}}
\caption{Multiplicities of photons and neutron hits on the proton-remnant side expected from the EPICS and DPMJET-III models for p-Pb collisions at 4.4 TeV center of mass energy in the proton remnant side.}
\label{fig.multiplicity}
\end{center}
\end{figure}

To experimentally investigate these important physics aspects, the LHCf collaboration has proposed to the LHCC to take data during the next proton-Lead run at LHC, currently foreseen for the end of 2012. We have submitted to LHCC a Letter of Intent ~\cite{LOIpPb}, detailing the physics case for this type of measurement, and showing that the existing detector configuration (Arm2 in particular) is  well suited for this purpose. Some interesting results are shown in the last part of this section, based on the simulations performed using EPICS and DPMJET-III hadronic interaction models to describe the proton-Lead collisions. A total of 10$^7$ events have been generated for each model, considering a 3.5 TeV proton hitting a 1.38 TeV/nucleon Lead ion with a null crossing angle and random impact parameter. The total energy in the nucleon-nucleon center of mass frame is approximately 4.4 TeV. The most significant results are obtained by measuring particles produced in the proton remnant side, that is more simply accessible from the experimental point of view; in this region smaller multiplicity of neutral particles are expected, as can be seen from Fig.~\ref{fig.multiplicity}, that reports the expected number of photons and neutrons hitting both the Arm2 towers. 

Approximately 1\% of the events have one single hit, and less than a fraction of 10$^{-5}$ of the events have two hits on a single tower. Therefore we do not expect particular problems both during data taking and analysis phases. Additionally, we want to stress that the absolute energy scale can be directly checked also in the p-Pb run with the data by the direct $\pi^0$ reconstruction method, by separately identifying the two photons of the $\pi^0$ decay in the two separate towers, as has already been demonstrated with the 7 TeV pp data~\cite{Menjo:2011zz}.

\begin{figure}
\begin{center}
\resizebox{0.9\columnwidth}{!}{\includegraphics{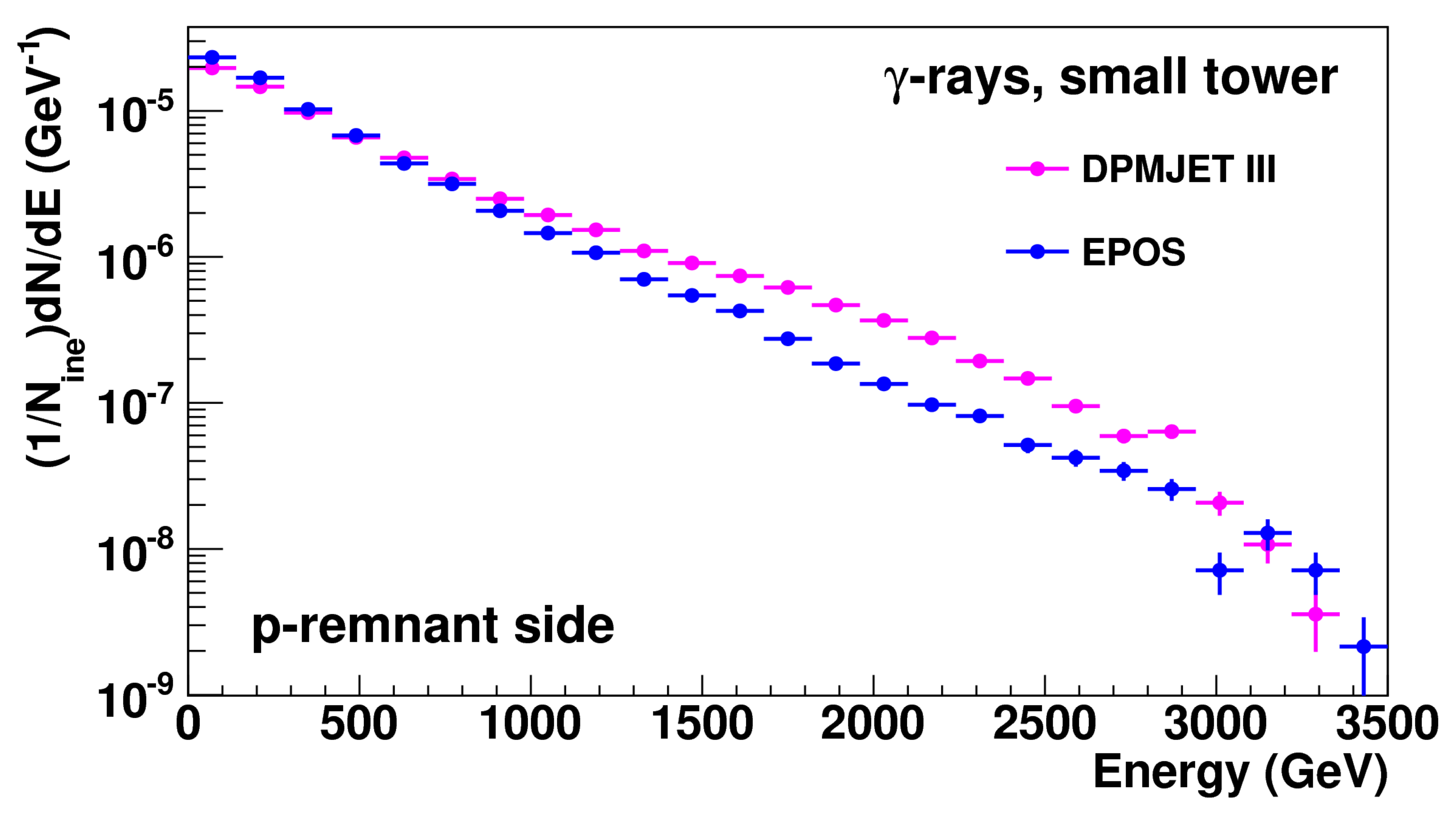}}
\resizebox{0.9\columnwidth}{!}{\includegraphics{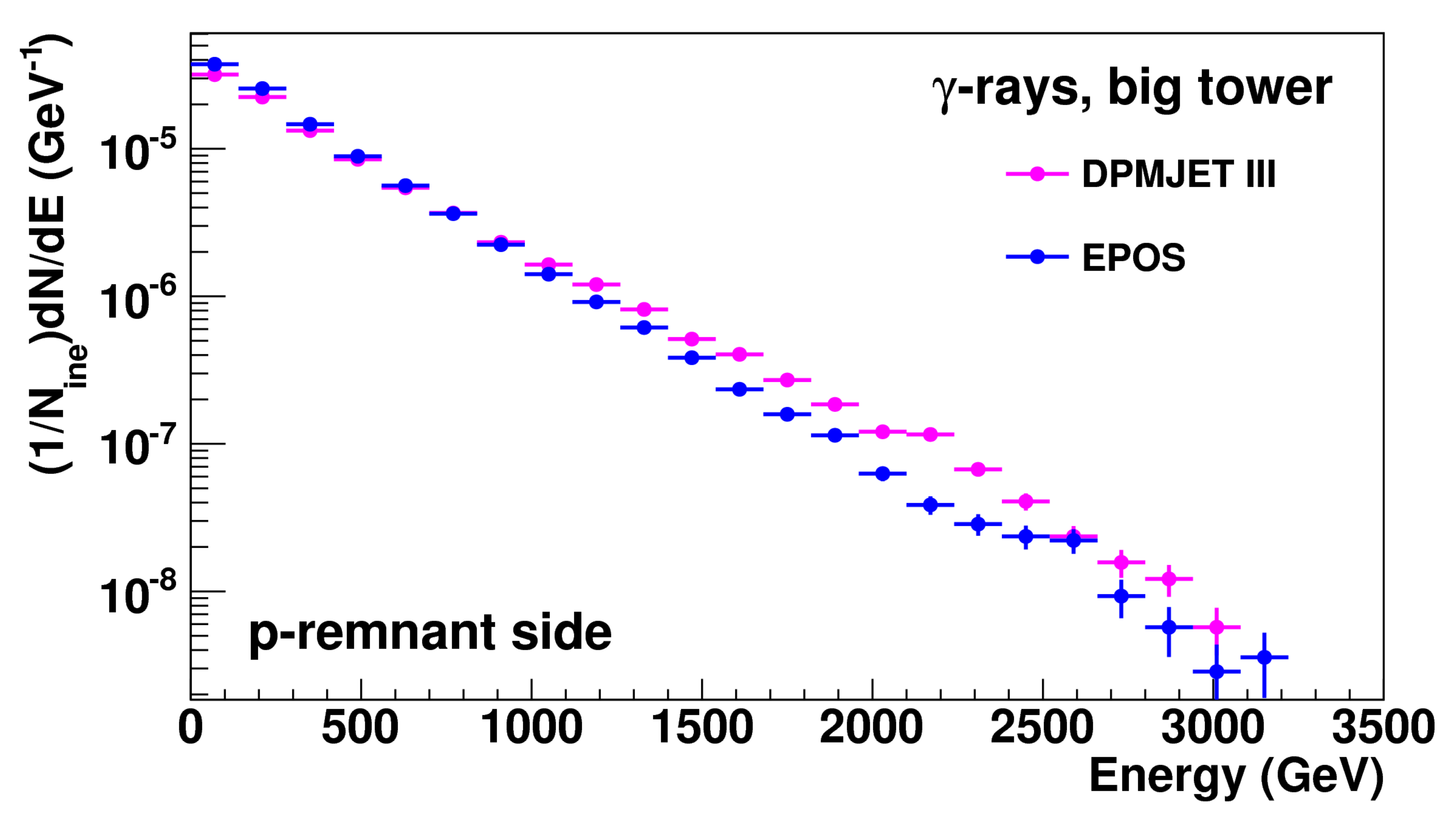}}
\caption{Expected energy spectra of photons hitting the Arm2 detector on the proton-remnant side small tower (top) and big tower (bottom). The statistics shown in the plots correspond to a data taking period of few hours at the nominal 10$^{28}$cm$^{-2}$s$^{-1}$ luminosity.}
\label{fig.spectra_pPb}
\end{center}
\end{figure}

Figure~\ref{fig.spectra_pPb} shows the expected photon spectra in the small (top) and large (bottom) tower, 
as obtained with the EPICS and DPMJET-III models, for 10$^7$ inelastic p-Pb collisions. 
A clear and significant difference between the two models can be seen from this figure. In particular we find that the DPMJET-III model has a harder spectra than the EPOS model. The difference in the slopes is evident and LHCf will easily discriminate between the two models. 

\section{Future activities}

\begin{figure}
\begin{center}
\resizebox{0.99\columnwidth}{!}{\includegraphics{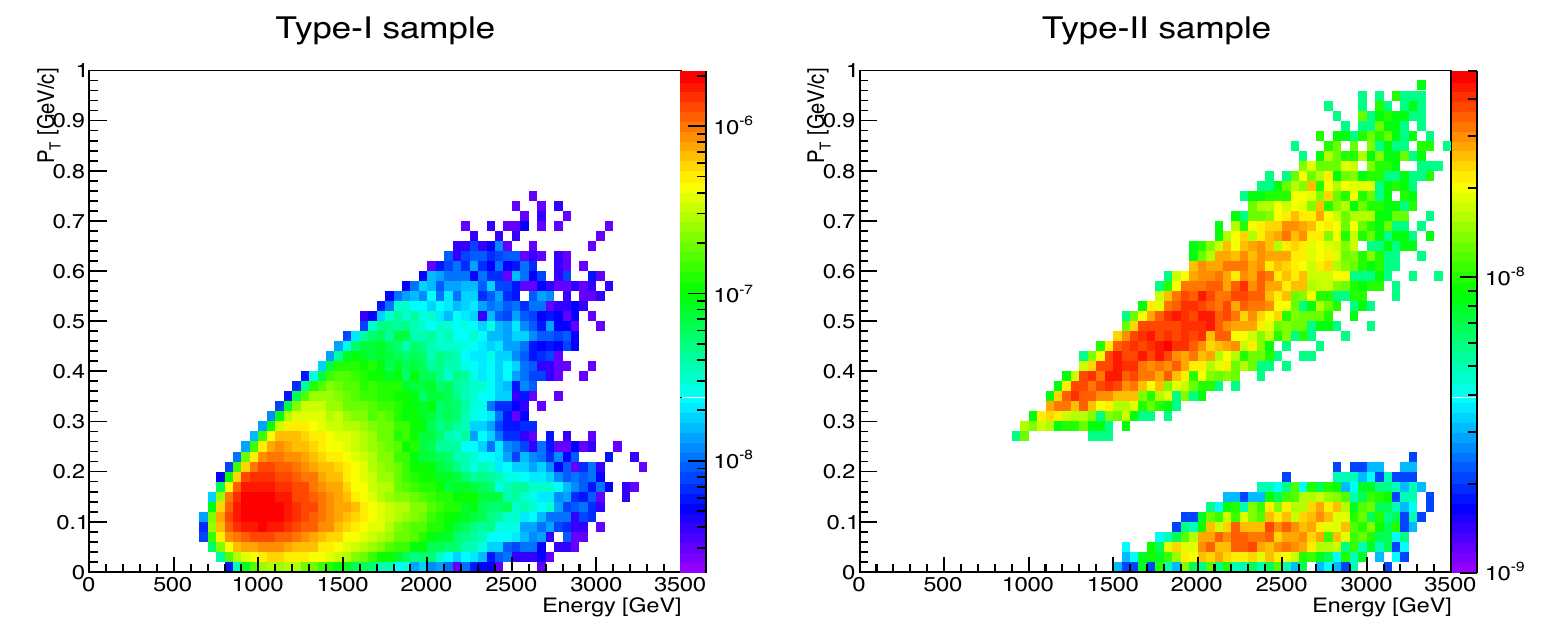}}
\caption{LHCf p$_T$ vs energy acceptance region covered by type I (left) and type II (rigth) $\pi^0$ reconstruction}
\label{fig.pi0}
\end{center}
\end{figure}

 New analyses of data collected in 2009-2010 runs are in progress, in particular the measurement of the single photon 
spectra at 900 GeV is almost finalised as well as the reconstruction of $\pi^0$ spectra. \\ 
The possibility to reconstruct also $\pi^0$ from pair of photons impinging in the same tower (type II) in addition to 
the $\pi^0$ reconstruction from pair photons impinging one in each tower (type I)
is under investigation. As can be seen from Fig.~\ref{fig.pi0} the reconstruction of type II $\pi^0$ events gives the 
possibility to explore a different p$_T$ region: while type I events are mainly characterised by pair of photons with 
large opening angle and dominates at low energies, type II $\pi^0$ events are characterized by tight photon opening angles 
and dominates at higher energies. The combination of type I and type II reconstruction  thus enlarges our physics acceptance.

In the meantime, the LHCf Collaboration is also working on the upgrade of the detector to improve the 
radiation resistance in view of the 14 TeV p-p run, currently foreseen in 2014. The scintillating part of the 
detector will be replaced with GSO slabs. Beam test results~\cite{gso} 
have demonstrated that the upgrade of the calorimeter with GSO will enable LHCf to sustain the radiation level foreseen 
in the 14 TeV run. 
Additional improvements in the front-end electronics of the silicon position sensitive layers of ARM2 detectors as well 
as an optimization of 
the layout to improve the stand-alone silicon energy resolution are on going.
\section{Summary}
LHCf has measured for the first time the single photon energy spectra in the very forward region of 7 TeV 
p-p collisions at LHC. None of the hadron interaction models mostly used in HECR Physics show a 
perfect agreement within the statistical and systematic errors with the data. New analyses and additional studies are 
in progress and the detector upgrade for the p-p run at 14 TeV is on going.


\begin{thebibliography}{}
\bibitem{Abreu:2011pj}  P.~Abreu, et al.  [The Pierre Auger Collaboration],
\emph{The Pierre Auger Observatory I: The Cosmic Ray Energy Spectrum and Related Measurements} arXiv:1107.4809 [astro-ph.HE].
\bibitem{Tsunesada:2011mp}
  Y.~Tsunesada [for the Telescope Array Collaboration],
  arXiv:1111.2507 [astro-ph.HE].
\bibitem{jinst}O. Adriani, et al., \emph{The LHCf detector at the CERN Large Hadron Collider}, \emph{JINST} {\bf 3} (2008) S08006. 
\bibitem{plb} O. Adriani, et al., \emph{Measurement of zero degree single photon energy spectra
                  for $sqrt(s) = 7$ TeV proton-proton collisions at LHC}, \emph{Phys. Lett.} {\bf B703} (2011) 128.
\bibitem{DPMJET} F.W. Bopp, et al., \emph{Antiparticle to Particle Production Ratios in Hadron-Hadron and d-Au Collisions in the DPMJET-III Monte Carlo}, \emph{Phys. Rev.} {\bf C77} (2008) 014904.
\bibitem{qgsjet} S. Ostapchenko, \emph{Monte Carlo treatment of hadronic interactions in enhanced Pomeron scheme: I. QGSJET-II model.}, \emph{Phys. Rev.} {\bf D83} (2011) 014108.
\bibitem{sybill} E.-J. Ahn, et al., \emph{Cosmic ray interaction event generator SIBYLL 2.1}, 
\emph{Phys. Rev.} {\bf D80} (2009) 094003.
\bibitem{epos} K. Werner, et al., \emph{The hadronic interaction model EPOS}, \emph{Nucl.Phys.Proc.Suppl.} 
{\bf 175-176} (2008) 81.
\bibitem{pythia} T. Sj\"{o}stand, et al., \emph{A Brief Introduction to PYTHIA 8.1}, \emph{Comput. Phys. Comm.} {\bf 178} (2008) 852.
\bibitem{auger} J.Bellido for the Auger Collaboration, \emph{The distribution of shower maxima of UHECR 
air-showers} in proceedings of \emph{32$^{st}$ ICRC, Beijing}, (2011).
\bibitem{RHIC} J. Adams, et al., \emph{Forward Neutral Pion Production in $p+p$ and $d+\mathrm{Au}$ Collisions at 
$\sqrt{{s}_{NN}}=200$ GeV}, Phys. Rev. Lett. 97 (2006) 152302.
\bibitem{LOIpPb} O. Adriani, et al., \emph{LHCf Letter of Intent for a p-Pb run. A precise study of
                      forward physics in $\sqrt{s_{NN}}= 4.4$ TeV proton-Lead ion
                      collisions with LHCf at the LHC}, CERN-LHCC-2011-015. LHCC-I-021 (2011).
\bibitem{Menjo:2011zz}
  H.~Menjo, {et al.}, \emph{Monte Carlo study of forward pi0 production spectra to be measured by the LHCf experiment for the purpose of benchmarking hadron interaction models at 10$^{17}$ eV}, Astropart.\ Phys.\  {\bf 34} (2011) 513.
\bibitem{gso} K. Kawade, et al., \emph{Study of Radiation Hardness of Gd$_2$SiO$_5$ scintillator for 
Heavy Ion Beam}, \emph{JINST} {\bf 6} (2011) T09004.
\end{thebibliography}
\end{document}